\documentclass[12pt]{article}

\usepackage{sbc-template}
\usepackage{graphicx,url}
\usepackage{amsmath}
\usepackage[utf8]{inputenc}  
\usepackage{placeins}
\usepackage{hyperref}
\usepackage[english]{babel}
\usepackage{listings}
\usepackage{color}
\definecolor{mygreen}{rgb}{0,0.6,0}
\definecolor{mygray}{rgb}{0.5,0.5,0.5}
\definecolor{mymauve}{rgb}{0.58,0,0.82}

\title{Skin color independent robust assessment of capillary refill time}

\author{Raquel P. de Souza Bachour; Eduardo Lopes Dias; George C. Cardoso}
\address{Departamento de Física, FFCLRP,\\
Universidade de São Paulo de Ribeirão Preto, 14040-901,Brazil}
\providecommand{\keywords}[1]{\textbf{\textit{Keywords---}} #1}
\begin{document}
\maketitle

\begin{abstract}
Capillary refill time (CRT) is a method for evaluating peripheral
	perfusion by visual assessment. CRT is especially useful
for quick evaluations in the absence of
	sophisticated equipment. However, there are repeatability and
	reproducibility limitations with CRT, especially for dark skin. To test the limits of CRT repeatability and skin color independence, we developed a system and method to perform simple and robust CRT measurements. The
	system consists of an RGB camera and an LED lamp, with
	crossed circular polarizers imaging to attenuate the light reflected by the
	superficial layer of the skin. The capillary refill time is
	determined using an exponential regression on the time-dependent green channel mean pixel
	intensity of the region of interest after the
	compression is released. We limited this regression up to a data-dependent cut-off time, after which we assume the exponential model is invalid, and used the confidence interval of the uncertainty to develop a criterion to flag and discard faulty measurements. We tested the system on twenty-two
	volunteers with skin phototypes ranging from I to VI on the
	Fitzpatrick scale, applying to their forearms a 7 kPa compression for 5 s. After the release of measurements flagged as inadequate (about 20\% of measurements) by our regressions, our results indicated good precision, with high repeatability for all skin phototypes. Approximately 80\% of measurements fall within $\pm 20\%$ of the individual's expected value for CRT (mean CRT value). Our results suggest CRT can be used as a quantitative measurement and encourages further developments for the implementation of a similar method on smartphone cameras for quick and robust CRT measurements in patients' triage, monitoring, and telehealth.
\end{abstract}


\section{Introduction}
Capillary refill time (CRT) is one of the most widely
acknowledged and used methods \cite{Schriger1988,Osborn2004} to
estimate peripheral perfusion status
\cite{Lima2009,ValenzuelaEspinoza2014, King2014}, for quick assessment or in low-resource environments. 
CRT is defined as the time required for a distal capillary bed to
regain its normal color after having received enough mechanical
compression to cause blanching \cite{Nickel, Pickard2011} of the
skin surface. Compression is typically applied by a finger of the person who measures,
who uses a chronometer and their own visual assessment to measure the refill time 
\cite{watson1993,Raju1999,Bridges2017}.
CRT measurements sites in humans include
the sternum \cite{Crook2013}, 
on the forearm \cite{Blaxter2015},
in the legs and feet \cite{Bridges2017,Gorelick1993,Ballaji2021}, in the fingertips \cite{Mrgan2014,Monteerarat2021},
and the knees \cite{Pandey2013,Ait-Oufella2014}. 
When executed in ideal conditions by trained professionals, 
CRT has been used to diagnose septic shock \cite{Otieno2004}, 
dehydration in children \cite{Lima2009,Shavit2006,Fleming2015}, and
viral diseases, such as dengue \cite{Dengue1}, 
and, more recently, as a
prognosis factor in COVID-19 patients \cite{Yormaz2021}.

Among CRT's main advantages are simple
equipment, high speed, and simplicity in training. 
Yet, CRT's adoption is hampered by concerns regarding inter and intra-observer reproducibility,
a lack of standardization for the pressure and for the
duration of the compression \cite{ValenzuelaEspinoza2014,Pickard2011,Shinozaki2019a,Anderson2008},
the effect of external factors such as the room's lighting \cite{Brown1994} and the temperature
of the limb and the environment \cite{Pickard2011,Gorelick1993,John2018}, and the
effect of skin color, particularly dark skin, on 
CRT accuracy \cite{Nickel,Ait-Oufella2014,Matas2001}.
These limitations have called into question the applicability and
usefulness of manual CRT measurements \cite{Pickard2011,Otieno2004,Lewin2008}.
Attempts to improve the
reliability and objectivity of CRT measurements, include the proposal of a device that utilizes
optical assessment of diffuse reflectance on the skin to
calculate the CRT \cite{Blaxter2015}, a device
comprising a compressible plastic optical fiber to measure the CRT under the foot \cite{Ballaji2021}, and a 
video camera system for training personnel for performing 
traditional CRT measurements \cite{Chang2017}.

Video-based CRT measurements \cite{Shinozaki2019a,Shinozaki2019,Kerr2018,CIZMECIOGLU2022,TollJohn2019} have also been proposed due to the greater sensitivity and linearity RGB cameras have when 
compared to the human eye. Cameras allow for the detection of subtle hues and intensity changes between the time of skin compression and its capillary refilling to the original state.
While some studies simply visually analyze the CRT videos at a later time \cite{TollJohn2019}, 
others automate the video processing \cite{Kerr2018}. 
Shinozaki et al. \cite{Shinozaki2019a} acquire
the RGB channels' intensities during a fingertip test, fit an exponential decay between the instants of maximum compression and of 90\% recovery, and obtain CRT with success. We could not find studies in the literature addressing issues such as CRT uncertainties, reproducibility, reliability, and robustness with respect to skin phototype.
In addition, the uncertainties of CRT under robust, controlled conditions, are not known so far.

In the present paper we show that CRT can be made robust and reliable, at least by using controlled compression, video processing, and polarized light. By robust, we mean insensitive to perturbations such as measurement repetitions, and measurement reproductions with changes of skin phototypes. 
By reliable, we mean that inadequate or poor measurements are identified and flagged to be repeated. Our method is based on recording a video of a region of interest (ROI) after the release of the
compression, and uses image processing and curve fitting to
calculate the CRT, and crossed circular polarizers
between the light source and the camera to attenuate the light reflected on the outer surface of the skin. 
We tested our method on 22 volunteers, including all Fitzpatrick phototypes,
under controlled conditions of temperature, lighting, and applied
pressure.  

To distinguish our CRT results from the literature ones, we name our capillary refill time \textit{pCRT} (\textit{polarized} CRT). The polarized light is not essential, but is known to increase robustness in skin measurements. This terminology will be adopted for the remainder of the paper.

\section{Materials and methods}
\subsection{Study subjects}
Twenty-two healthy volunteers (20–70 years), comprising all Fitzpatrick skin types (I–II; III-IV, and V-VI) and both sexes (9 female and 13 male) chose to participate in this study after being detailed about the procedure (University of São Paulo Ethics Committee CAAE $95342518.1.0000.5407,$ $3.046.098/$ FFCLRP). 
Volunteers were recruited in the campus community and in its extended network. We chose to invite volunteers that represented all Fitzpatrick skin types.

\subsection{Experimental protocol}
\begin{figure}[h]
	\centering
	\includegraphics[scale=0.5]{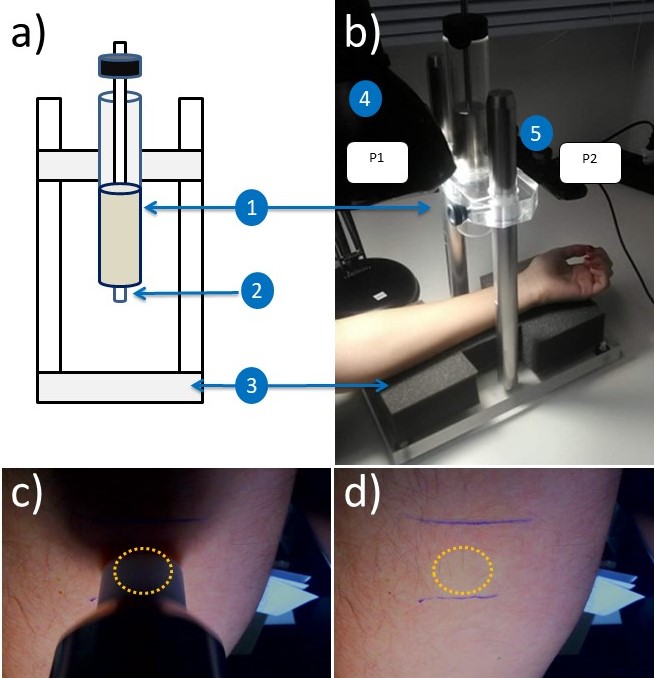}
	\caption{
		\textbf{Experimental setup.} 
		a) Schematic
		illustration of the weight and arm support, front view;
		(1) Standard aluminum cylindrical weight; (2) $ 4 cm^2$
		thermally insulating Teflon tip, that comes into contact
		with the subject's skin; (3) Armrest (dense polyurethane
		foam). 
		b) Setup with a volunteer's arm in actual measurement position; (4) light
		source and (5) video camera, with crossed circular
		polarizers (P1 and P2) installed on each. 
		c) Weight lowered on a volunteer's forearm as viewed by the video acquisition camera. The ROI is
		highlighted by the dotted circle.  
		d) Blanching of the ROI after release of compression.
		}
	\label{fig:fig1}
\end{figure}

We built a cylindrical compression device to produce skin
blanching on the volunteer’s forearm (Figure \ref{fig:fig1}(a)).
The cylindrical device was terminated by a 4 cm$^2$, rounded
Teflon contact surface, and smoothly slid inside a hollow
external acrylic cylindrical vest to be gently set to rest on the
volunteer’s forearm. The Teflon insulates the metal cylinder from
the skin to prevent heat transfer.

A light source (LED TKL 90 – 14 W, E27, Taschibra Ltda, Brazil) illuminated the forearm. The light source was turned on 15 minutes before the start of each acquisition, for stabilization. A video camera (HD Pro-C910, 24 fps, resolution of 1280 × 720 pixels, Logitech S.A., Switzerland) and the light source were attached to the fixture (Figure \ref{fig:fig1} (b)). The camera was focused on each subject's region of interest (ROI) on the forearm, 9 cm from the wrist line. Circular polarizers ($\lambda/4$ = 125 nm, 99.98$\%$ polarization efficiency, crossed transmission 0.5$\%$, 3D Lens Corp., Taiwan ROC) installed in front of the light source and camera lens (P1 and P2 in Figure \ref{fig:fig1} (b)) with crossed polarizations, attenuated light reflected on the outer layer of the skin \cite{Jacques2002,Groner1999}. 

The experiments were performed in a temperature-controlled room (20$^{\circ}$C – 22$^{\circ}$C) as suggested by Pickard et al. \cite{Pickard2011}. All videos were acquired in a dark room illuminated only by the circularly polarized light source.
Before the start of the measurements, the volunteers remained
seated for 10 minutes for acclimatization. Volunteers sat in a
relaxed position on a height-adjustable chair with their left arm
positioned approximately at heart level (Fig \ref{fig:fig1} (b)),
and monitored by an oximeter (CMS50D—USB, ROC) attached to
their left index finger. For each measurement, the camera started
recording the ROI for 10 s before the weight was lowered on the
subject's left forearm, where it remained for 5 s applying a
pressure of 7 kPa, after which it was lifted and the capillary
refill phenomenon was recorded. The recording was stopped 20 s 
after the lifting of the weight, which is much longer than the capillary refill times.
These pCRT measurements were repeated five times for each subject, with
a 1-minute rest between measurements.

\subsection{Video analysis and pCRT calculation}
\begin{figure}[h]
	\centering
	\includegraphics[width=13cm]{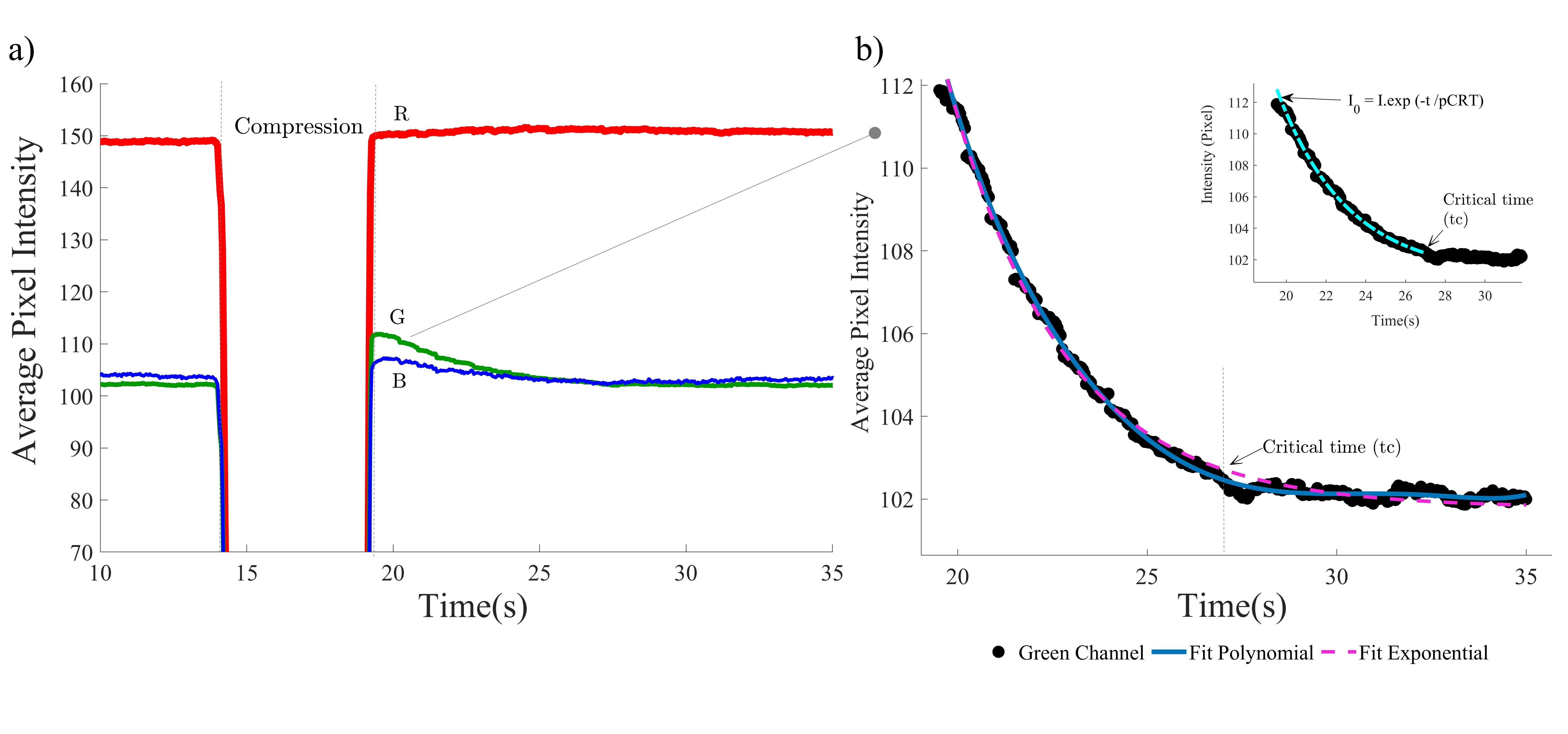}
	\caption{\textbf{Mean ROI pixel intensities during a pCRT experiment.}
		a) Mean intensities of the R, G and B channels from
		the pixels inside the ROI. The
		cylindrical weight blocking the camera during compression causes the sharp drop in
		intensities observed between 14 s and 19 s. After the weight is lifted, the G channel 
		displays a pronounced peak and a
		decay, which is highlighted in b) Behavior of the G channel after
		the compression is lifted. The 
		decay is approximately exponential (dashed line) but levels off after a
		 \textit{cut-off time} ($t_c$); the inset shows the proper exponential decay region (two-dashed line) used to determine pCRT.
	}
	\label{fig:fig2}
\end{figure}

\begin{figure}[h]
	\centering
	\includegraphics[scale=0.5]{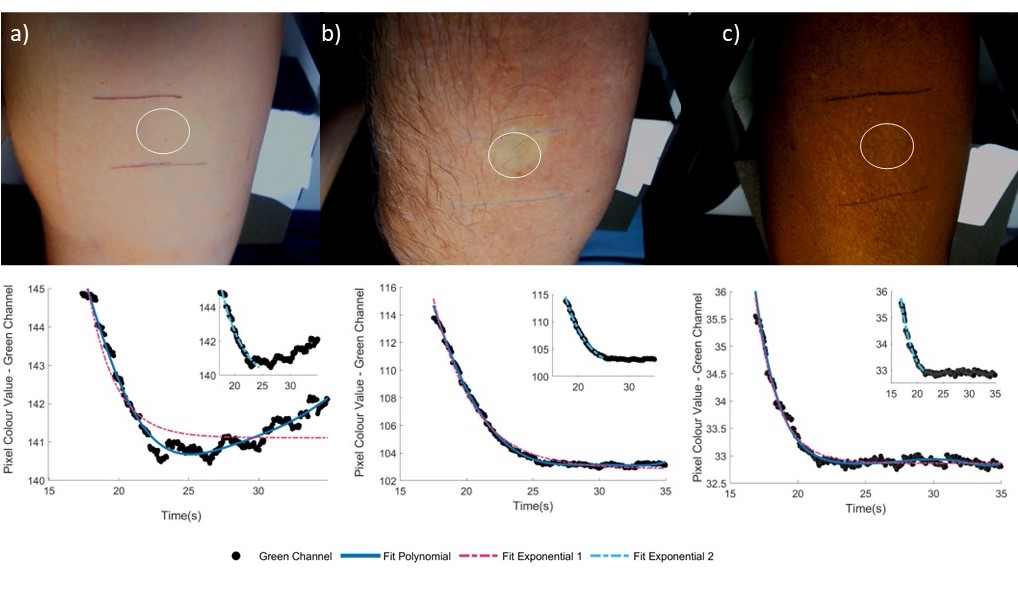}
	\caption{
		\textbf{Forearm of three volunteers
		of different phototypes} imaged immediately after removal of the
		7 kPa compression on the ROI (marked
		by a circle). a) Phototype I-II, b) Phototype III-IV
		and c) Phototype V-VI. The corresponding G-channel mean ROI intensity decay and curve regressions are shown below each volunteer image. Notice that curve behavior is not exponential for times longer than $t_c$. 
	}
	\label{fig:fig3}
\end{figure}

We analyzed the videos (110 videos, 5 for each of the 22 volunteers) with our pCRT calculation routines implemented in Matlab version 2015a (MathWorks, MA, USA). The release of the cylindrical weight from the skin surface
causes a pronounced color change. The average intensities of the R, G and B channels of the ROI pixels are calculated for each frame (Figure \ref{fig:fig2} (a)) and the green (G) channel (Fig. \ref{fig:fig2}(a)) presents the best signal-to-noise ratio. 
We hypothesized that the behavior of this channel can be modeled by an
exponential decay curve. However, just fitting an exponential function
from the time of the G channel's peak to the end of the recording
proves ineffective, as the channel's intensities become noisy
after some time and the agreement with an exponential model
quickly fails (See Figure \ref{fig:fig2}(a)) and Figure \ref{fig:fig3}). Thus, we have devised a multi-step protocol to realize the exponential regression. 

First, we identified the cut-off time $t_c$ after which the exponential model significantly diverges from the observed curve.
To find, $t_c$ we simultaneously fit a sixth-order polynomial and a provisional exponential decay on the entire G channel intensity after the release of compression. 
The polynomial regression is used as a low pass filter and for interpolation of the data. The instant $t_c$ represents the instant of maximum difference between 
the polynomial and the provisional exponential
function (Figure \ref{fig:fig2}(b)) and Figure \ref{fig:fig3}(a)). This procedure proved to be robust for all our instances. Finally, pCRT
is the time constant of yet another exponential decay function:

\begin{equation}
	I=I_o exp{(-t/pCRT)}.
	\label{eq:1}
\end{equation}

That is adjusted to the original data only between the maximum value of the green channel, and $t_c$ (inset of Fig. \ref{fig:fig2}(b)). We also calculate the 95\% confidence interval (CI) $\sigma_{95\% CI}$ (an uncertainty), in the regression of equation \ref{eq:1} to the data. The horizontal and vertical offsets in equation \ref{eq:1} have been omitted for simplicity.
This whole procedure was applied to every video that was acquired.

\subsection{Statistics and reproducibility test}
We adopted the dimensionless quantity, $\sigma _{95\% CI}/pCRT$ which henceforth we refer to as the \textit{coefficient
of variation}, as the main metric for the uncertainty of our
results. This relative regression uncertainty simplifies comparison between different pCRT measurements.

The reproducibility of pCRT was evaluated by analyzing the
distribution of measurements from each participant, from phototype groups, and for all subjects together. We also sought to establish a method to choose a maximum acceptable value of the relative regression uncertainty for a single measurement, which we call the
\textit{discard-threshold}, to flag and remove incorrect measurements, while keeping plausible ones. 

\section{Results}

In healthy tissue, after the skin is bleached out by compression, the color returns rapidly as the blood refills the dermal capillaries. This color return is the foundation of the CRT test. Our pCRT method calculates capillary refill time by analyzing the ROI's image intensity over time.
As shown in Figure \ref{fig:fig2}, the exponential decay of intensity characteristic of capillary refill is most clearly distinguishable in the green channel (G). The lower signal-to-noise (SNR) ratio of the G channel held for all our
measurements, across all subjects. Thus, we chose to perform our analysis on the G channel only.

Table \ref{table:2} summarizes pCRT results for each skin phototype group. Notice that the mean pCRT and
its uncertainty in this table were calculated for all
participants of each phototype group, regardless of age. CRT is known to increase with age \cite{Anderson2008a}. The pCRT values do not differ significantly despite the different phototypes, which is suggest pCRT to be robust with respect to light absorption by melanin. Differences in the mean pCRT might stem from differences in the age groups and respective standard deviations, but the small number of volunteers in each group prevents further interpretation.

\begin{table}[h]
	\centering
	\caption{
		Mean pCRT for different Fitzpatrick skin types. SD = standard deviation.}
    \begin{tabular}{ c c c c } 
    		\hline 
    		&pCRT$\pm$SD (s)  
    		& Age$\pm$SD (yr.)
    		&Number of volunteers\\
    		\hline
    		Phototypes I—II	 & 4.0 $\pm$ 0.7 & 27 $\pm$ 12  & 8\\
    		Phototypes III—IV &	4.4 $\pm$ 1.3 & 46 $\pm$ 14  & 9\\
    		Phototypes V—VI &	3.7 $\pm$ 1.7 & 44 $\pm$ 19  &5\\
    		\hline
    \end{tabular}
	\label{table:2}
\end{table}

We have observed that in some volunteers with white skin (phototypes I-II), and in some volunteers with 
dark skin (phototypes V-VI), the color change due to the
capillary refill phenomenon was difficult or impossible to observe with the
naked eye, or even visually on the recorded video. 

\begin{figure}[h!]
	\centering
	\includegraphics[scale=0.5]{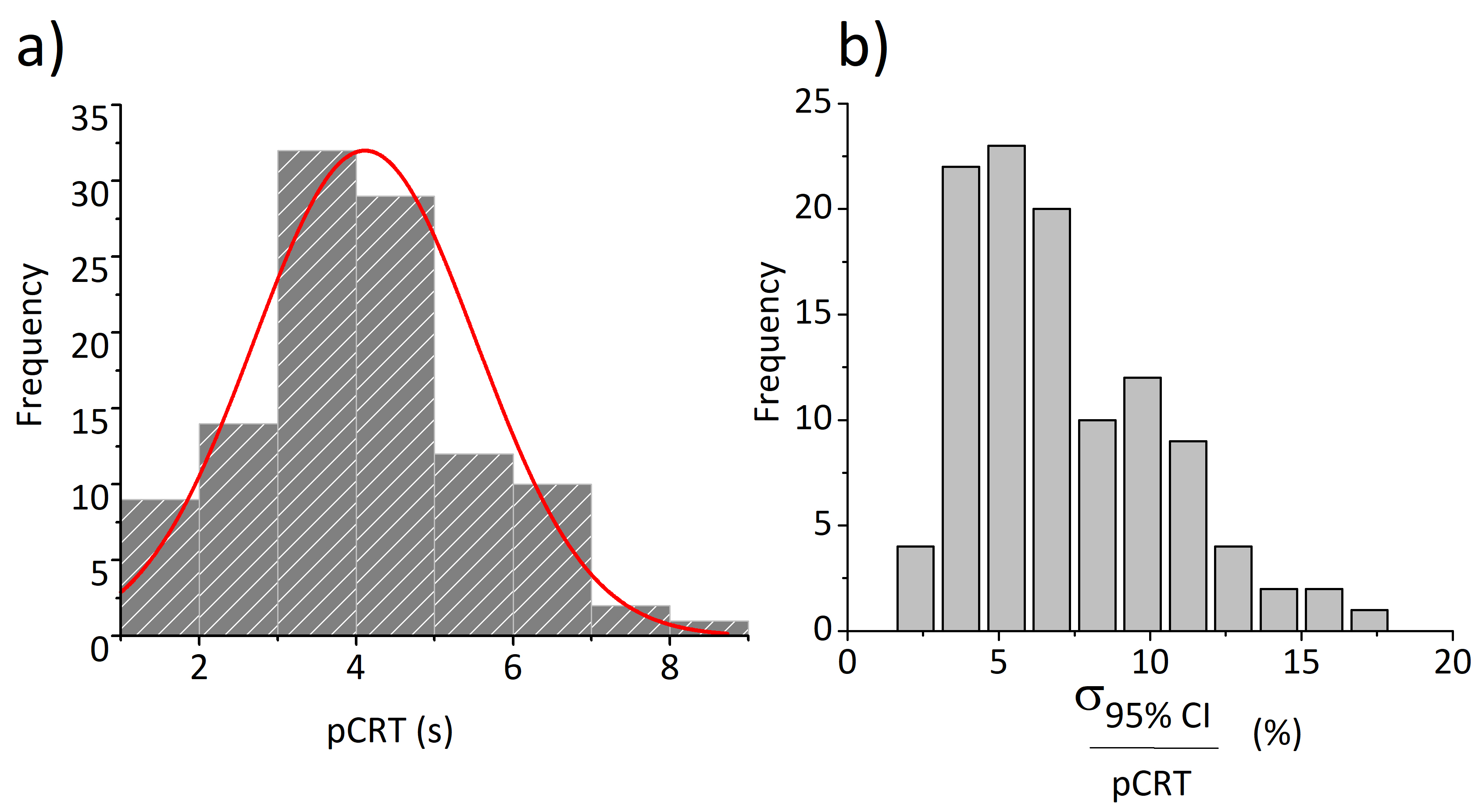}
	\caption{
		Descriptive statistics for all pCRT readings. a) The distribution of pCRT results for all 110 data
		points (5 measurements for each volunteer). The red line
		is a Gaussian fit (mean =3.9; standard deviation =1.3). b) Frequency distribution of he coefficient
		of variation $\sigma_{95\%CI}/pCRT$ (mean = 7.1\%).
	}
	\label{fig:fig4}
\end{figure}

Figure \ref{fig:fig4}(a) displays unique pCRT measurements with mean 3.9 s, and Figure \ref{fig:fig4}(b) displays the respective relative regression uncertainties \\ $\sigma_{95\% CI}/pCRT$ with mean $7.1\%$. 
One measurement showed $\sigma_{95\% CI}/pCRT$
greater than $20\%$, which we omitted in
Figure \ref{fig:fig4}, in the calculation of the means.

\begin{figure}[h]
	\centering
	\includegraphics[scale=0.45]{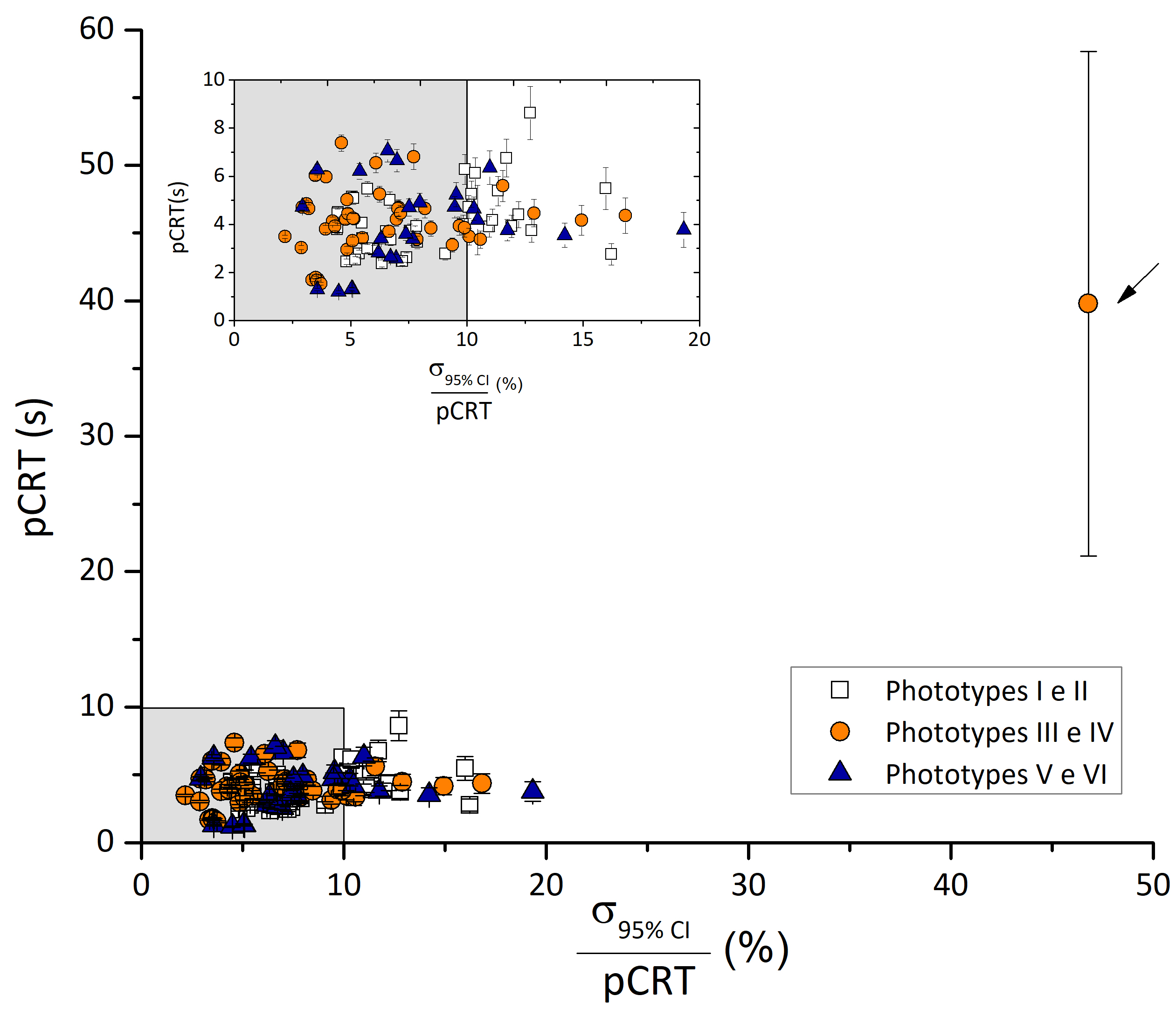}
	\caption{
		\textbf{ Distribution of the pCRT reading for all phototype subgroups.} All 110 readings are shown (5 for each of the 22 subjects). The error bars represent $\sigma_{95\% CI}$.  
		The arrow highlights a point with abnormally high regression uncertainty, 
		which indicates an erroneous measurement.
		The gray area in the inset shows the region with relative regression uncertainty below 10\%, which
		is our readings discard-threshold. Note that
		all phototypes are approximately equally represented and
		evenly distributed inside the gray box.}
	\label{fig:fig5}
\end{figure}

\begin{figure}[h]
	\centering
	\includegraphics[scale=0.45]{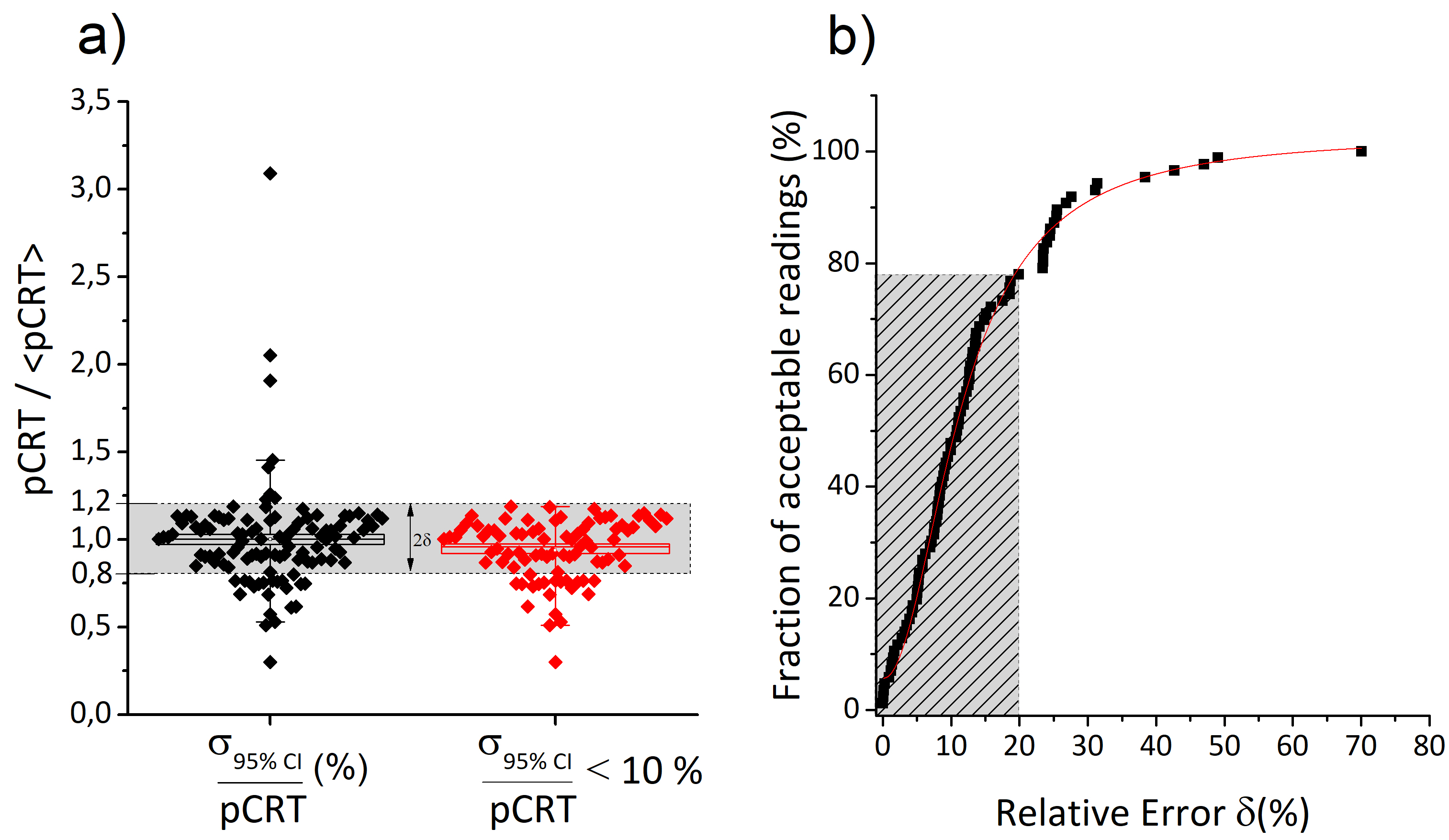}
	\caption{\textbf{ Repeatability study.} 
		a) Each point is a reading (110 in total). The vertical axis is the ratio between each
		measurement and the subject's expected (mean) pCRT calculated
		from 5 repetitions of readings. The horizontal axis is
		qualitative (Boxplot grouping of data). The
		distribution on the left (black diamonds) is before
		the application of the discard-threshold (110 points),
		and on the right (red diamonds) is after the application of the discard-threshold (86 points).
		b) Fraction of acceptable measurements as a function of the Relative error (difference between a given measurement and the average of the 5 measurements). The highlighted region (gray box pattern) shows that for a $\delta = 20\%$ maximum relative error, for example, the fraction of acceptable readings is $78\%$. The red line shows logistic fit with $R^2 = 0.995$.
	}
	\label{fig:fig6}
\end{figure}

Figure \ref{fig:fig5} illustrates the uncertainty
distribution of the measurements, detailing the three different phototype subgroups.
Most measurements have a relative regression uncertainty below $10\%$, which is
about 0.4 s. Notice the evident outlier has a relative regression uncertainty greater than $45\%$.
Thus, we decided to analyze the measurement performance, discarding results with $\sigma_{95\% CI}/pCRT$ greater than $10\%$ (right side of Figure\ref{fig:fig6}(a)). 

The vertical axis of Figure\ref{fig:fig6}(a) represents the value of pCRT readings from curve regressions normalized by the mean pCRT ($<pCRT>$) obtained from the 5 measurements for each individual. If repeatability were perfect, $pCRT/<pCRT> = 1$ for all data points, independently of the individual's $<pCRT>$, and the relative error $\delta = pCRT/<pCRT> - 1$ for all pCRT readings relative to the (“true”) mean $<pCRT>$ of its own set of 5 measurements would be zero.

For a given discard-threshold (we chose $10\%$) we can build a curve to diagnose the metric that resembles the receiver operating characteristic (ROC) curve. Such curve shows how the fraction of acceptable readings increases as the acceptable relative error $\delta$ increases (Figure\ref{fig:fig6}(b)). To build such ROC curve, we integrate the fraction of single measurements as a function of relative errors from zero to the maximum error. When high relative errors are not acceptable, only a small fraction of all measurements will be acceptable.

Our chosen discard-threshold entailed that approximately 80\% of the original 110 readings remained, and out of these 80\%
of measurements on a given subject fell within $<pCRT> \pm 20\%$ (marked in gray in Figure\ref{fig:fig6}(b)),
where $<pCRT>$ is the subject's average pCRT calculated from the
5 total measurements. We can also see that $95\%$ of the readings have a relative error lower than $35\%$. As an aside, we can see that Figure\ref{fig:fig6}(b) is well fitted by a logistic curve ($R^2 =0.995)$, which indicates the error distribution is essentially Gaussian even after application of the discard-threshold.  

A stricter discard-threshold (lower), flags and rejects more pCRT readings the ROC curve would rise faster, which means that a larger fraction of the remaining readings would have a low relative error. A compromise must be made between discarding and repeating readings, and simply averaging multiple readings or risking an incorrect result.

\section{Discussion}

We have verified that measuring peripheral perfusion status by capillary refill time can be done when in a repeatable and robust way, with low pressure applied. For that end, we used a polarized light, controlled pressure, a video camera, and image processing. The use of crossed circular polarizers attenuates the reflection component of light captured by the digital camera, enabling visualization of deeper regions of the skin \cite{Stockford2003, Borovkova2019}. Our system produced successful measurements in subjects with dark skin (phototypes V and VI), which are a challenge for visual CRT measuring method \cite{Matas2001, Saavedra1991, Bickler2005}, and has not been demonstrated by other studies.

We are not aware of earlier studies that performed the CRT test in volunteers of all skin phototypes (Table \ref{table:2}). For all volunteers studied, the behavior curve of the green channel’s average intensities indicates a better signal-to-noise ratio (SNR) compared to the other channels.
Our study also developed a flagging recipe to reject most poor readings. Readings flagged as inadequate are discarded and need to be repeated by the operator. The discard-threshold is adjustable and a lower probability of an erroneous measurement can be reached, at the cost of increasing the fraction of rejected readings (Figure\ref{fig:fig6}). 
In our setup, approximately 80\% (CI 95\%) of the readings were within 20\% of the expected pCRT, when a discard-threshold of 10\% relative regression uncertainty was adopted. Averaging of 2 or 3 measurements would further reduce the chance of error above 20\% in the estimated pCRT.

The return of skin color after compression by the cylinder (Figure\ref{fig:fig3}), depends on the state of hydration and elasticity of the tissue\cite{Popov2017}. After a certain time of decay (the steep part of the decay) in some individuals, the curve becomes noisy. We believe that this noise relates to mechanical changes in elasticity of the skin, maybe caused by a different time dynamics of subjacent fat or muscle (Figure\ref{fig:fig3}, graphs below the images). Thus, to stabilize the exponential regression, we established a cut-off time to delimit the region where the exponential fit is valid. This cut-off time varies from reading to reading and corresponds to the time when a 6th order polynomial used as a low-pass filter for the data, and an auxiliary exponential decay function, diverge the most from each other when fitted to the G channel intensity average. The cut-off strategy improves the quality of the exponential regressions and the repeatability of the readings. Except for the cut-off strategy, our regression method follows approximately the one proposed by Shinozaki et al.\cite{Shinozaki2019} that calculates CRT fitting and exponential decay to the grayscale video signal. Shinozaki et al. use a cut-off between 90\% and 10\% of the decay curve, and do not take advantage of regression uncertainties.

Though pCRT relates to the same physiological parameters as CRT and yields values similar to the visual CRT measurement method \cite{Pickard2011,Champion1980}, we must note that the two quantities are not exactly equivalent. The capillary refill times calculated through our method are, in general, longer than visual CRT. The difference may be also due to the lower pressure we apply (7 kPa) compared to conventional CRT \cite{Blaxter2015,Shinozaki2019,Kawaguchi2019, Liu2020}. The compression pressure applied to induce whitening of the ROI is one of the many factors known to influence the CRT \cite{John2018, Tuchin2022}. Ordinarily, these compressions are subjective and are typically applied with the examiner’s fingertip. Different researchers have proposed different compressions. For example, Kawaguchi et al. propose a pressure of 10 kPa - 70 kPa applied with the fingertip for 2 seconds as optimal \cite{Kawaguchi2019}. Other studies have proposed 17 kPa \cite{Blaxter2015, Liu2020}, and 60 kPa \cite{Shinozaki2019}. We have adopted throughout this study the lowest pressure yet, 7 kPa, which is low enough not to induce any pain in the forearm. With this low pressure, we demonstrated repeatability.
In another study to be published elsewhere, we noticed that application of high pressure in the forearm (23kPa) increased noise, decreased fit quality, and repeatability. Our success with using low pressure (7 kPa) may be attributed not only to the higher sensitivity of digital cameras but also to the use of crossed circular polarizers, which improves SNR by attenuating the component due to reflection on the skin surface \cite{Groner1999,Stockford2003, Borovkova2019}. We believe that with adequate image processing for removal of reflection, the polarizers will be unnecessary.

Limitation of this study include possible interference of cardiovascular and parasympathetic systems of the volunteers during the five CRT readings. Volunteers may have found the experiment to be stressful or at least initially uncomfortable due in part to the cold, unlit environment, unfamiliar equipment and the requirement to stay still during most of the process. This situation may have caused the activation of the sympathetic nervous system of some participants during data acquisition, which induces a change in the heart rate \cite{Schriger1988,Cho2019}. Heart rate and temperature are factors known to influence CRT \cite{Jacquet-Lagreze2019}, this may have caused intra-participant pCRT variation along the 5 measurements. Other limitations of this study are the lack of skin temperature measurement, and the relatively small number of volunteers did not allow for an investigation of how pCRT varies with HR. 

The robustness to different phototypes and good repeatability of pCRT opens up the possibility for health condition status tracking and physiological monitoring studies where the conventional CRT method has proved unreliable. Among possibilities that remain to be studied are the relationship between pCRT and temperature, heart rate, blood pressure, or with the autonomic nervous system.

\section{Conclusion}
Our primary contributions to the field is to show that CRT can be made robust,
observer- and skin-color independent, and can be performed at
low compression, using simple equipment. A development based on our pCRT methodology is promising for further
research in a clinical setting. It may potentially be used to reliably assess the capillary refill time of an individual as function of time for health condition tracking.


\section{Acknowledgements}
The authors thank Beatriz Janke and Carlos Renato da Silva, for experimental help, and all the volunteers who participated in the experiment. 

\section{Supplementary Materials}
The code used for this work is available at https://github.com/Photobiomedical-Instrumentation-Group/pCRTMatlab

\section{Author Contributions}
Conceptualization R.P.d.S.B., E.L.D, and G.C.C.; 
Formal analysis, R.P.d.S.B., E.L.D, and G.C.C; 
Methodology, R.P.d.S.B.;
Software, R.P.d.S.B.; 
Supervision, G.C.C.; 
Validation, G.C.C.; 
Writing—original draft, R.P.d.S.B.;
Writing—review and editing, R.P.d.S.B., E.L.D, and G.C.C.
All authors contributed to the article and approved the submitted version.


\section{Funding}
This study was financed by the Coordenação de Aperfeiçoamento de Pessoal de Nível Superior—Brazil (CAPES)—Finance Code 001.

\section{Informed Consent Statement}
Informed consent was obtained from all subjects involved in the study.

\section{Conflict of interest}
The authors declare no conflicts of interest.

\bibliography{referencias}
\end{document}